\documentclass[authoryear, review]{elsarticle}
\usepackage[labelfont=bf,justification=raggedright,singlelinecheck=false]{caption}
\usepackage{lineno,hyperref}
\modulolinenumbers[5]

\journal{Dynamics of Atmospheres and Oceans}

\usepackage{verbatim}
\usepackage{graphicx}
\usepackage{amsmath}
\usepackage{amssymb}

\graphicspath{{Figures/}}

\newcommand{\<}   {\langle}
\renewcommand{\>} {\rangle}

\newcommand{\bs}  [1]{\boldsymbol{#1}}

\renewcommand{\d}  {\mathrm{d}}
\newcommand{\dd}  [2]{\frac{\mathrm{d} #1}{\mathrm{d} #2}}

\begin{document}

\begin{frontmatter}

\title{The effect of surface buoyancy gradients on oceanic Rossby wave propagation}

\author{\textsc{Xiao Xiao}}

\author{\textsc{K. Shafer Smith}}
\address{Courant Institute of Mathematical Sciences, New York University, New York, NY}

\author{\textsc{Shane R. Keating}}
\address{School of Mathematics and Statistics, University of New South Wales, Sydney, NSW 2052, Australia}

\begin{abstract}

  Motivated by the discrepancy between satellite observations of
  coherent westward propagating surface features and Rossby wave theory,
  this paper revisits the planetary wave propagation problem, taking
  into account the effects of lateral buoyancy gradients at the
  ocean's surface.  The standard theory for long baroclinic Rossby
  waves is based on an expansion of the quasigeostrophic stretching
  operator in normal modes, $\phi_n(z)$, satisfying a Neumann boundary
  condition at the surface, $\phi_n'(0) = 0$.  Buoyancy gradients are,
  by thermal wind balance, proportional to the vertical derivative of
  the streamfunction, thus such modes are unable to represent
  ubiquitous lateral buoyancy gradients in the ocean's mixed layer.

  Here, we re-derive the wave propagation problem in terms of an
  expansion in a recently-developed ``surface-aware'' (SA) basis that
  can account for buoyancy anomalies at the ocean's surface. The
  problem is studied in the context of an idealized Charney-like
  baroclinic wave problem set in an oceanic context, where a surface
  mean buoyancy gradient interacts with a constant interior potential
  vorticity gradient that results from both $\beta$ and the
  curvature of the mean shear.  The wave frequencies, growth rates and
  phases are systematically compared to those computed from a
  two-layer model, a truncated expansion in standard baroclinic modes
  and to a high-vertical resolution calculation that represents the
  true solution.  The full solution generally shows faster wave
  propagation when lateral surface gradients are present.  Moreover,
  the wave problem in the SA basis best captures the full solution,
  even with just a two or three modes.

\end{abstract}


\end{frontmatter}

\linenumbers

\section{Introduction}

Satellite altimetric observations show that wave speeds in the ocean
are systematically greater than those predicted for linear first
baroclinic Rossby waves \citep{chelton96}. Several mechanisms have
been suggest to explain these ``too-fast'' westward propagating
surface signals: Doppler shifting and alteration of the PV gradient by
the background mean flow \citep{Killworth97b}; topographic decoupling
of upper-ocean waves \citep{Tailleux01}; the combination of the two
effects \citep{Killworth03a}; conflating wave propagation with the
westward propagation of coherent mesoscale eddies
\citep{Chelton2011,Early2011}.  Our aim here is neither to contradict
nor promote the relevance of these approaches, but rather to point out
the effect on surface signal propagation speeds of yet another nearly ubiquitous
characteristic of the ocean: lateral gradients of surface buoyancy.

The possibility that surface gradients might significantly affect wave
propagation is also suggested by their known effects on eddies.  Eddy
stirring against surface buoyancy gradients effectively generates
ample surface buoyancy anomalies, and a large number of recent studies
indicate surface mesoscale and submesoscale structures consistent with
the effects of such anomalies on quasigeostrophic dynamics
\citep[e.g.][]{Xu2011,Ponte13,Wang13} .  The limiting case of a flow
entirely controlled by the surface buoyancy field is referred to as
``surface quasigeostrophic'' (SQG) theory.  As an example for how
surface gradients can affect wave propagation, consider the extreme
case of vanishing interior potential vorticity (PV) gradients. The
resulting Rossby edge wave has phase speed $\sim 1/ \kappa$, where
$\kappa ^2 = k^2 + l^2$ and $(k,l)$ is the two dimensional wavenumber.
By contrast, the speed of a linear first baroclinic Rossby wave,
derived by only assuming background interior PV gradients, is $\sim 1/
\kappa^2$.

\begin{figure}
\includegraphics[scale=1]{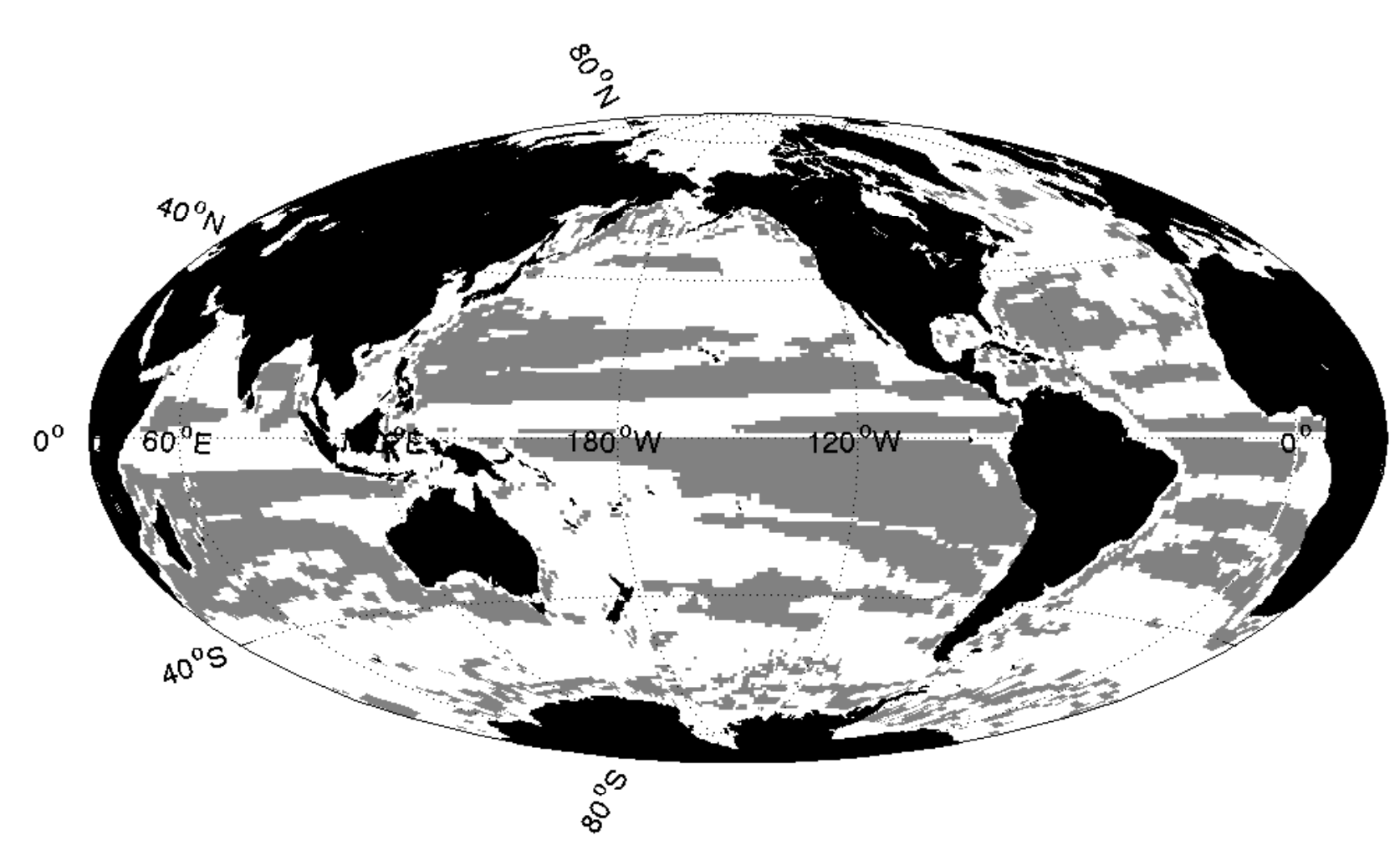}
\caption{Global maps of Charney instability (gray). OCCA (OCean
  Comprehensible Atlas) annal mean data is used \citep{Forget2010}. From
  the instability criteria, the ratio of $dU/dz$ at ocean surface and
  the mean meridional PV gradient, $Q_y$, are computed and averaged
  over the top 500 meters. Charney type instability occurs when this
  ratio is negative, which is shown in gray.}
\label{drhody}
\end{figure}

In most places, of course, the oceanic mean state exhibits both
surface buoyancy and interior potential vorticity gradients;
moreover, the mean state is almost always baroclinically unstable
\citep{Tulloch2011}. Cases where instabilities are caused by
interactions between surface shear flows and interior PV gradients are
analogous to those of the classic atmospheric Charney instability
problem \citep{Pedlosky87}; following \cite{Smith07b}, we refer to
these as ``Charney type'' instabilities.  Fig.~\ref{drhody} shows an
estimate of geographical regions that are Charney-unstable (see
caption for details). In these regions, the existence of surface
buoyancy gradients will change the nature of both the instability and
the waves;  and even in regions that are stable to Charney-type
baroclinic instability, the surface buoyancy gradient may affect the wave
propagation characteristics.  Since it is difficult to separate propagation of a wave
through the changing of mean state of the ocean from the production of
baroclinically unstable waves from the mean state itself, we address
this by focusing on the dispersion relation at small wavenumber, where
the flow is stable or nearly stable.

Here we shall consider an idealized ``ocean-Charney'' problem,
constructed by demanding the mean zonal velocity has a non-zero shear
at the upper surface, zero shear at the lower surface, and a constant
PV gradient in the interior;  for example, with constant buoyancy
frequency $N$, the resulting mean flow is quadratic in the vertical
coordinate. [To add the effects of interior PV gradients, this can be
augmented by the addition of a mean flow component proportional to the
first baroclinic mode (a cosine in the case of constant $N$).]  A
numerical solution of the resulting eigenvalue problem for horizontal
plane waves indeed shows that surface buoyancy gradients yield faster waves,
relative to equivalent cases with no surface gradients (see
Fig.~\ref{fullxi}; see caption and next section for details).

In addition to considering the physics of Rossby waves in the presence
of surface buoyancy gradients, we also seek to find an efficient and transparent
model for this process.  Most of our theoretical understanding of wave
and eddy dynamics in the ocean is based on models that have been
simplified in the vertical: either by modal truncation or by using a
small number of isopycnal layers. For example, \citet{Flierl78}
studied the equivalence between layered models and continuous
stratified flows and proposed a two-mode model that overcomes the
inaccuracy due to the density step and upper-layer thickness of a
layered model.
Here we propose a model based on a truncated expansion in a set of
surface aware (SA) modes recently developed by \citet[][SV13 hereafter]{Smith13}.  The SA modes represent
surface-interior dynamics in a natural way, efficiently representing
surface intensified motions, driven by surface buoyancy gradients, with just a few
modes. The model is systematically compared to layered and standard
modal truncations, and used to explore the effects of surface lateral
buoyancy gradients on the dispersion relation.

The goals of this study are
\begin{enumerate}
\item to explore Rossby wave propagation on a
  background mean flow with surface buoyancy gradients, and
\item to introduce a new truncated model that efficiently captures the
  effect of surface buoyancy gradients on the wave problem.
\end{enumerate}
The paper is organized as follows. Section \ref{oceancharney} builds
the linear ocean Charney plane wave problem and gives a brief review
of the derivation of linear plane wave solutions for the
quasigeostrophic (QG) equation.  Full numerical solutions for the
problem are computed and discussed.  In section \ref{2sa}, we
introduce the SA modes of SV13, and derive a wave model based on a
truncated set of these modes.  Section \ref{solutions} presents
analytic solutions for the case of constant stratification with
certain mean flows, as well as numerical solutions for a broader range
of flows.  In addition, these solutions are compared to solutions of a
two-standard-mode truncated model, as well as of the classic two-layer
model.  The efficiency of surface-aware modes is also
discussed. Finally, section 5 concludes the paper.

\section{A linear ocean Charney plane wave model}\label{oceancharney}

\subsection{Construction of the background flow}

We shall consider the quasigeostrophic (QG) plane wave problem linearized about an
idealized zonal mean flow $U(z)$, with corresponding mean interior
meridional PV gradient $Q_y(z)$ and mean meridional upper surface ($z=0$)
buoyancy gradient $B_y$ given by
\begin{equation}\label{meangrad}
  Q_y(z) = \beta - \dd{}{z}\left(\frac{f_0^2}{N^2}\dd{U}{z}\right)
  \quad\text{and}\quad
  B_y = -f_0 \dd{U}{z}(0),
\end{equation}
respectively, where $f_0$ is the mean Coriolis parameter, $\beta$ is
the meridional Coriolis gradient, and $N(z)$ is the buoyancy
frequency.  The lateral buoyancy gradient at the lower surface $z=-H$
is taken to vanish.  The zonal background flows $U(z)$ are chosen to
have the properties
\[
\dd{U}{z}(0) = \Lambda^\text{dim}, \quad 
\dd{U}{z}(-H)= 0, 
\quad\text{and}\quad
\frac{1}{H}\int_{-H}^0U(z)~\d z = 0,
\]
where the superscript ``dim'' is added to distinguish the dimensional
upper surface shear from a nondimensional version defined below.
To proceed the velocity is written as a sum of two parts,
\[
U(z) = U^S(z) + U^I(z),
\] 
where $U^S(z)$ is the ``surface'' part, with $\d U^S/\d z = \Lambda^\text{dim}$
at $z=0$ and vanishing lower surface shear, and $U^I(z)$ is the
``interior'' part, with vanishing shears at both the upper and lower
surfaces.  Ideally the interior PV gradient would be entirely
controlled by the interior part of the mean shear, but the best we can
do is demand the contribution of $U^S(z)$ to the PV gradient be a
constant.  The above constraints on the surface derivatives then imply
\begin{equation}\label{surfu}
  \dd{}{z}\left(\frac{f_0^2}{N^2} \dd{U^S}{z}\right) 
  = \frac{f_0^2\Lambda^\text{dim}}{HN_0^2},
\end{equation}  
where $N_0 \equiv N(0)$.

A convenient choice for the interior part, satisfying the above
constraints, is $U^I = U_1\Phi_1(z)$, where $\Phi_1(z)$ is the gravest
non-constant eigenfunction (i.e. the first baroclinic mode) for the
standard vertical mode problem
\begin{equation}\label{standardmodes}
  \dd{}{z}\left(\frac{f_0^2}{N^2} \dd{\Phi_j}{z} \right) = -\lambda_j^2 \Phi_j, 
\quad\text{with} \quad
\dd {\Phi_j}{z} = 0 \quad\text{at}\quad z=0, -H,
\end{equation}
and $\lambda_j$ is the inverse Rossby radius for mode $j$.  For
reasons explained below, \textit{we normalize $\Phi_1$ such that its
  minimum value is -1}.  The resulting composite flow $U(z)$ has thus
an associated PV gradient consisting of a constant part plus a
component proportional to the first baroclinic mode,
\begin{equation}\label{meanpv}
Q_y(z) = \beta  - \frac{\Lambda^\text{dim} f_0^2 }{H N_0^2} + U_1\lambda_1^2\Phi_1(z).
\end{equation}

\subsubsection*{Nondimensionalization}

The parameters controlling the structure of the PV and surface
buoyancy gradients are $\beta$, $\Lambda^\text{dim}$ and $U_1$.  To reduce the
parameter space, and keeping in mind that we wish to compare results
here to those of standard Rossby wave theory, we nondimensionalize
lengths by the deformation scale $L_R \equiv N_0H/f_0$, and speeds by
the long Rossby wave speed $U_R \equiv \beta L_R^2$. From here
forward, all variables should be taken as nondimensional. The
nondimensional surface buoyancy gradient 
is
\[
\frac{B_y}{f_0U_R/H} = -\frac{f_0^2 \Lambda^\text{dim}}{\beta H N_0^2}
\equiv -\Lambda
\]
and the nondimensional interior PV gradient is then
\[
\Pi(z) \equiv \frac{Q_y(z)}{\beta} = 1 - \Lambda + \xi \Phi_1(z)
\quad\text{where}\quad
\xi \equiv \frac{U_1\lambda_1^2}{\beta}.
\]

Baroclinic instability in this flow is possible when either $\Lambda$
has the opposite sign as $\Pi$ for some $z$ (Charney-type
instability), or when $\Pi$ itself changes sign (Phillips-type
instability), or both.  If $\Lambda=0$, then by our choice of
normalization that min$(\Phi_1)=-1$, there will be a sign change in
$\Pi$ if $\xi>1$, hence $\xi$ is the supercriticality parameter for
Phillips-type baroclinic instability.  If $\xi=0$ and $\Lambda\neq 0$,
then $\Lambda$ and $Q_y=1-\Lambda$ must have the opposite sign, which
will be true for $\Lambda<0$ and $\Lambda>1$ (thus the flow is stable
only when $0<\Lambda<1$).  When both parameters are nonzero, one can
get stable flows, or instabilities of the Phillips, Charney or mixed
type, with regime boundaries along the lines $\xi = \pm (1-\Lambda)$.

\subsubsection*{Special Case I: Uniform background stratification}

For constant background stratification $N(z) = N_0$, the
nondimensional surface component of the mean velocity is
\begin{equation}\label{usconstant}
U^S(z) = \Lambda\left(\frac{z^2}{2}+ z + \frac{1}{3}\right).
\end{equation}
The first baroclinic mode with constant $N$ is $\Phi_1(z) =
\cos(\pi z/H)$, with $\lambda_1 = \pi/L_R$, thus the
interior part has the form
\[
U^I = \frac{\xi}{\pi^2}\cos(\pi z)\quad\text{with}\quad
\xi =\frac{U_1\pi^2f_0^2}{\beta H N_0^2}.
\]
The top right panel of Fig.~\ref{backn} shows $U^S$ and the baroclinic
modes $\Phi_j(z)$ are shown in the bottom left panel.

\subsubsection*{Special Case II: Nonuniform background stratification}

\begin{figure}[!htb]
\includegraphics[scale=0.9]{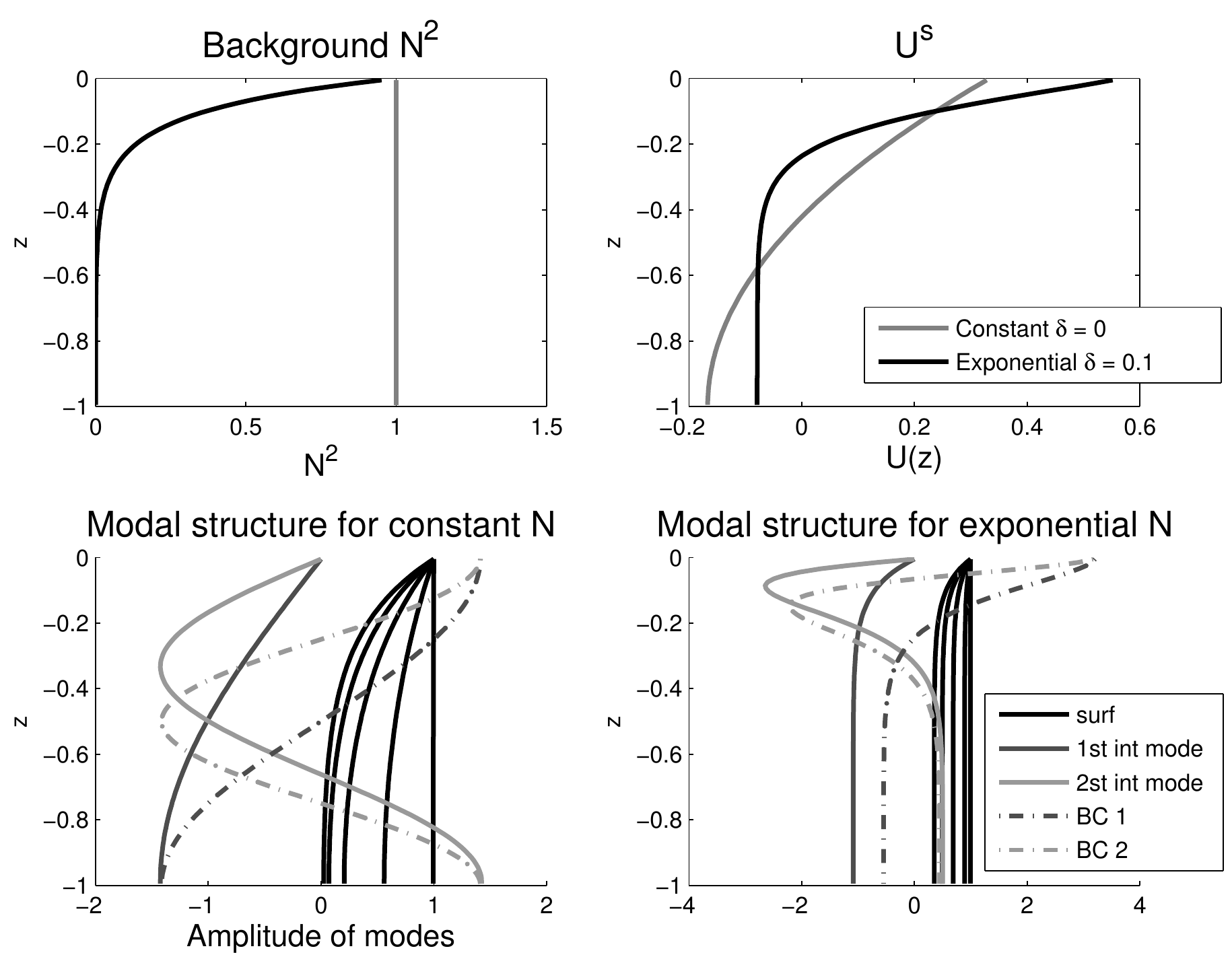}
\caption{Top left/right: Structure of background stratification and
  surface velocity, $U^S$, for constant/exponential stratification;
  Bottom left: Comparison of surface mode (from SA-mode basis) (solid)
  with barotropic (BT) mode (from standard normal modes basis)
  (dashed); Bottom right: Comparison of first two SA modes (solid)
  with first two standard modes (dashed).}
\label{backn}
\end{figure}

As an idealization of the oceanic thermocline, we also consider
\begin{equation}\label{Nexp}
N^2 = N_0^2 e^{z/\delta},
\end{equation}
where $N_0^2=N^2(0)$ and $\delta$ is the fractional scale depth. For
this particular choice of the stratification profile, the
non-dimensional surface velocity is, by equation
\eqref{surfu},
\begin{equation}
U^S = \Lambda\delta \left[e^{z/\delta} (z+1-\delta) 
+2\delta^2(1-e^{-1/\delta})-\delta\right].
\end{equation}
The interior flow $U^I$ is proportional to the first standard
baroclinic mode for the stratification \eqref{Nexp}, and $\xi$ depends
on its first eigenvalue.  Both can be computed from a WKB
approximation, but here we instead resort to a numerical solution.
The surface part of the mean flow is plotted in the top right panel of
Fig.~\ref{backn}, and the first baroclinic mode, $\Phi_1(z)$ --- to
which $U^I(z)$ is proportional --- is shown in the bottom right panel.

\subsection{Linear wave equations}

For a plane wave solution $\psi(x,y,z,t) =
\Re\{\psi_{kl}(z)\exp(kx + ly - \omega t)\}$ (and likewise for the
PV and surface buoyancy), the QG equations linearized about $U(z)$ are 
\citep[e.g.][]{Vallis06}
\begin{subequations}\label{qbk}
\begin{align}
(U-c) b_{kl} - \Lambda\psi_{kl} &=0, \qquad z = 0, \label{bk}\\
(U-c) q_{kl} + \Pi\psi_{kl} &= 0, \qquad -1 < z < 0 \label{qk}
\end{align}
\end{subequations}
where $c = \omega/k$ is the zonal phase speed, $\omega$ is the
frequency, and $q_{kl}(z)$ and $b_{kl}(z)$ are the complex
perturbation PV and surface buoyancy wave amplitudes at lateral wavenumber $(k,l)$,
respectively.  The (nondimensional) PV and surface buoyancy amplitudes are related
to the streamfunction amplitude $\psi_{kl}(z)$ by
\begin{equation}\label{fqpsi}
q_{kl} = -\kappa^2\psi_{kl} +( s \psi_{kl}')', 
\quad\text{and}\quad
b_{kl} = \psi_{kl}'(0)
\end{equation}
where $\kappa^2 = k^2 + l^2$ and $s(z) = N_0^2/N^2(z)$, and the prime
denotes a derivative in $z$.  From here forward, except where
confusion might ensue, the subscripts $kl$ will be dropped.

\subsection{Numerical solution to full wave problem}

\begin{figure}[!htb]
\includegraphics[scale = 0.9]{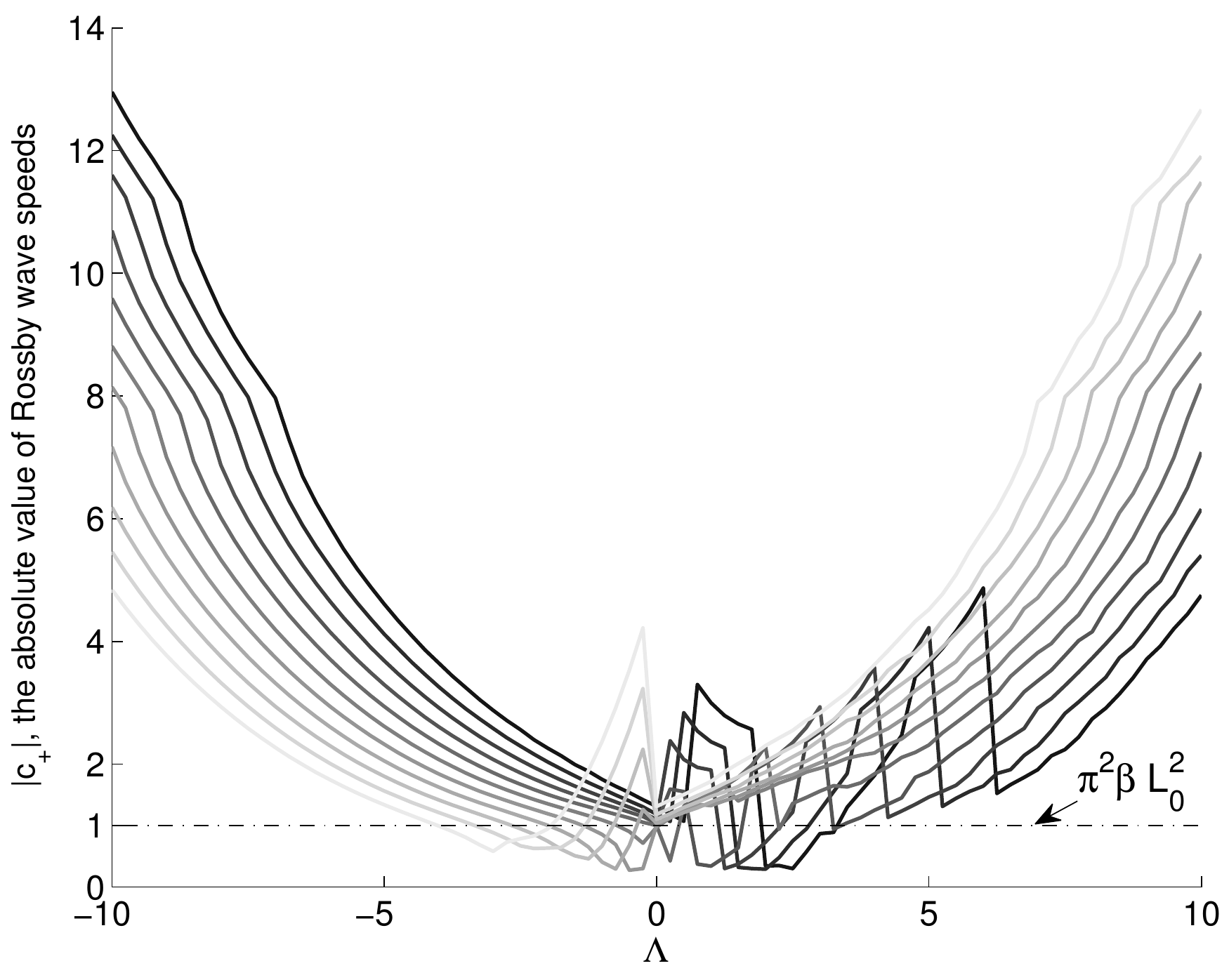}
\caption{Baroclinic Rossby wave speeds $c_+$ as a function of the
  upper surface buoyancy gradient $-\Lambda$ and the interior supercriticality $\xi$
  at wavenumber $\kappa=0.5$ (twice the deformation scale). The values
  of $\xi$ are integers varying from -5 (darkest) to 5 (lightest).
  Using OCCA data as in Fig.~\ref{drhody}, and assuming $\beta =
  2\times 10^{-11} ($ms$)^{-1}$, $f_0 = 10^{-4}$ s$^{-1}$, $N_0^2 = 4\times
  10^{-5}$ s$^{-2}$, and $H = 4000$ m, the nondimensional
  $|\Lambda|\lesssim 2$ .  In this nondimensionalization, the first
  baroclinic Rossby wave speed, $\beta (\pi L_D)^2$, is 1.}
\label{fullxi}
\end{figure}

Before proceeding to the truncated solutions, we present numerical
solutions of \eqref{fqpsi} for a range values of $\Lambda$ and $\xi$,
for the case $N=N_0$.  In particular, we adopt a finite difference
scheme with 100 equally spaced vertical layers, and solve the
resulting linear eigenvalue problem using MATLAB. The baroclinic
Rossby wave speed, denoted as $c_+$ hereafter, is the eigenvalue
corresponding to the second gravest mode in the full calculation.
Fig.~\ref{fullxi} shows baroclinic wave speeds, at wavenumber
$\kappa=0.5$, as functions of $\Lambda$, for a range of
$\xi$. According to our nondimensionalization, the baroclinic Rossby
wave speed for a resting mean state ($\Lambda = \xi = 0$) is 1
(denoted by the dashed line).  From the figure one can conclude that
in most places the presence of a surface buoyancy gradient speeds up the wave: the
magnitude of Rossby wave speeds will increase with increasing
$|\Lambda|$. The sharp jumps near $\Lambda=0$ occur when the type of
the stability of the system changes or the system is only weakly
unstable. For example, based on the instability criteria stated in the
last paragraph, when $\xi < -1$, as $\Lambda$ decreases from a large
positive value, the system switches from a Charney instability with an
eastward sheared surface velocity, to a mixed instability at $\xi =
1-\Lambda$, then again to a Phillips instability when $\xi$ crosses 0,
and finally to a Charney instability (with westward sheared surface
velocity) at $\xi \ge -1+\Lambda$.

\section{Modal projection on surface-aware modes}
\label{2sa}

Having shown that surface gradients do tend to speed up baroclinic
Rossby waves, we now attempt to capture this effect with a vertically
truncated model. The new model will be compared to traditional
truncations:  a two-layer model and a two-standard-mode model.

We first consider a truncation of the dynamics projected onto the
``surface-aware'' (SA) modes of SV13.  The reader is referred to that
paper for full details, and a review, specialized to the present case
with no buoyancy anomalies at the lower surface, is presented in
Appendix A.  Briefly, these modes are constructed to efficiently
capture both surface and interior dynamics, while diagonalizing the
energy.  Nondimensional weights $\alpha_\pm$ control how sensitive the
modes are to surface dynamics at the upper ($+$) and lower ($-$)
surfaces, and in the limit $\alpha_\pm\rightarrow\infty$
(corresponding to vanishing buoyancy anomalies at both surfaces), the
standard baroclinic modes are recovered.  In the ``ocean limit''
considered here,$\alpha_- \rightarrow \infty$, and $\alpha_+$ is
suitably small (see below), consistent with a lower surface that has
no significant buoyancy anomalies, but an upper surface that is dominated
by them.  In this limit, the gravest mode is a wavenumber-dependent
evanescent mode akin to the vertical structure of an SQG solution; this
mode becomes barotropic at very small wavenumber, and replaces the
standard barotropic mode.  The higher modes are oscillatory, like the
standard baroclinic modes, but shifted in phase.

Specifically, the SA streamfunction modes $\phi_j(z)$ in the ocean
limit are solutions to the eigenvalue problem
\begin{subequations}\label{samodes}
\begin{align}
\label{surfmodesa}
(s \phi_0')' &= \kappa^2 \phi_0, \quad\text{with}\quad 
\phi_0'(0)  = \frac{\mu_0^2}{\alpha_+}\phi_0(0),
\quad  \phi_0'(-1) =0\\
\label{intmodesa}
(s \phi_j')' &= -\lambda_j^2 \phi_j, \quad\text{with}\quad 
\phi_j(0) = 0, \quad \phi_j'(-1)  = 0, \quad j\ge 1
\end{align}
\end{subequations}
where $\mu_j^2 = \lambda_j^2+\kappa^2$, and it is assumed that
$\mu_0^2/\alpha_+ = O(1)$ and $\alpha_+ \ll 1$.  As explained above,
$\phi_0$ is a wavenumber-dependent evanescent `surface' mode and the
rest are oscillatory `interior' modes.  In addition, the modes are
orthogonal in the sense that
\begin{equation}\label{sanorm1}
\int_{-1}^0 s \phi'_i\phi'_j+\kappa^2\phi_i\phi_j~\d z = \mu_j^2\delta_{ij}.
\end{equation}
Note that other normalizations are possible (see SV13), but this one
is most convenient for the analysis presented here.  The traditional
baroclinic modes \eqref{standardmodes} are recovered in the limit
$\alpha_+ \to \infty$, and in this case $\phi_0$ becomes the
wavenumber-independent barotropic mode and the rest modes become
baroclinic, i.e. the boundary conditions for all modes become
$\phi_j'(0) = 0$.

For constant background stratification, the eigenvalue problem in
\eqref{samodes} can be solved analytically with normalization
\eqref{sanorm1}, with solutions
\begin{equation}\label{saconstN}
\phi_0 = \frac{\sqrt{\alpha_+}}{\cosh\kappa}\cosh\left[\kappa(z+1)\right] 
\quad\text{and}\quad
\phi_j = \sqrt{2}\sin\left[ \left(j-\frac{1}{2}\right)\pi z \right] + O(\alpha_+)
\end{equation}
for the surface and interior modes, respectively, with eigenvalues
\begin{equation}\label{saevalconstN}
\mu_0^2 = \alpha_+\kappa\tanh\kappa
\quad\text{and}\quad
\mu_j^2 = \kappa^2+(j-1/2)^2\pi^2.
\end{equation}
The structures of these ocean SA modes are shown in the bottom left
panel of Fig.~\ref{backn}. Notice that, as with an SQG solution, the surface mode
$\phi_0$ is wavenumber-dependent but with different $\kappa$
dependence than the SQG solution. The interior modes are sines instead
of cosines, as the traditional baroclinic modes are, which represent
interior motions with non-vanishing upper-surface derivatives.  For
the exponential stratification introduced in the previous section,
these modes are also surface intensified and the structures are
plotted in the bottom right panel of Fig.~\ref{backn}. In this case,
the surface mode is nearly barotropic except near the surface, and the
interior modes exhibit a similar structure to the traditional
baroclinic modes, except for the Neumann boundary condition.

Projecting the streamfunction wave amplitude onto a finite set of ocean SA
modes and using the orthogonality condition, the wave equation \eqref{fqpsi}
can be posed as a discrete eigenvalue problem
\begin{equation}\label{samatrix}
c\mathbf{a} = \mathsf{B} \mathbf{a}
\end{equation}
where $\mathbf{a} = \begin{pmatrix}
a_0, \dots,  a_n
\end{pmatrix} ^T$ is the coefficient vector and 
\begin{equation}\label{Bmatrix}
  \mathsf{B}_{ij} = \frac{1}{\mu_i^2}
  \left[\int_{-1}^0 \left(\mu_j^2U-\Pi\right)\phi_i\phi_j  ~ \d z 
   +  U(0)\phi_i(0)\phi'_j(0)-\Lambda\phi_i(0)\phi_j(0)
  \right]
\end{equation}
The phase speeds $c$ are the eigenvalues of the matrix $\mathsf{B}$;
the derivation of \eqref{samatrix} and \eqref{Bmatrix} are given in
Appendix A.

\section{Analytical and numerical solutions}\label{solutions}

In this section, we seek analytical and numerical solutions for the
ocean Charney plane wave problem proposed in section
\ref{oceancharney} with the truncated SA mode model discussed in
section \ref{2sa}. For cases with no interior flow ($U^I=0$), and either
constant or exponential stratification, we approximate the solutions
by using two-mode truncations in SA modes, and compare these to a
two-mode truncation using traditional baroclinic modes, as well as to
the wave speeds in a two-layer system. For the case when both
exponential stratification and interior mean flow are present, we use
a three-mode truncation for both modal bases to capture this more
complicated dynamical structure.

\subsection{Constant stratification, no `interior' flow}
\label{constantn}

Truncating the projected wave equation \eqref{samatrix} to just two
modes (i.e. $n=1$) results in a 2 $\times$ 2 matrix equation that can
be solved analytically if the integrals necessary to compute
$\mathsf{B}$ in \eqref{Bmatrix} can be computed in closed form.  For
constant stratification $N=N_0$ ($s=1$), the ocean-limit SA modes and
their eigenvalues are given in \eqref{saconstN}, and the necessary
integrals can be found in terms of standard functions; details are
given in Appendix B.  For our purposes here, we additionally neglect
the interior velocity, i.e. $\xi= 0$, so the interior mean PV gradient
is $\Pi = 1 -\Lambda$. Since there can be no sign change of the
interior PV gradient in this model, the only possible instability is
of the Charney-type, which may occur when $\Lambda$ and $\Pi$ have
opposite signs, e.g. when $\Lambda < 0$ or $\Lambda > 1$.

The details of the analytical solutions for the two-mode expansions in
both the ocean limit SA modes and the traditional modes, as well as for the
classic two-layer model, are relegated to Appendix B.  Here we
summarize just the small-$\kappa$ Taylor-expansions of the baroclinic
phase speeds for each of the three approximate solutions;
they are 
\begin{subequations}\label{smallkappa}
\begin{align}
\label{sacp}
\text{SA modes:}\quad 
c_+ &= -0.08 -0.02\Lambda   
&+ \left( 0.009 + 0.01\Lambda - 0.02 \Lambda^2\right)\kappa^2 
\quad+ O(\kappa^4),\\
\label{stcp}
\text{Standard modes:}\quad 
c_+ &= -0.025 + 0.13\Lambda  
&+ \left(0.01 - 0.01\Lambda + 0.02\frac{\Lambda^2}{1-\Lambda} \right)\kappa^2
\quad+ O(\kappa^2),\\
\label{2lcp}
\text{Two-layer:}\quad 
c_+ &= -0.125 &+\left(0.016-0.016\Lambda ^2\right)\kappa^2 
\quad+O(\kappa^4),
\end{align}
\end{subequations}
In the context of the standard baroclinic modal basis, $c_+$ is the
first baroclinic wave speed. Comparing the small-$\kappa$
approximations in \eqref{smallkappa}, one can see that both modal
truncated solutions depend on $\Lambda$ at $O(1)$, while the two-layer
approximation does not, thus the two-layer solution is not sensitive
to changes of surface buoyancy gradient strength at large scale. Note also that in
the standard-mode truncation \eqref{stcp}, there is a singularity at
$\Lambda=1$.

\begin{figure}[!htb]
\includegraphics[scale=0.9]{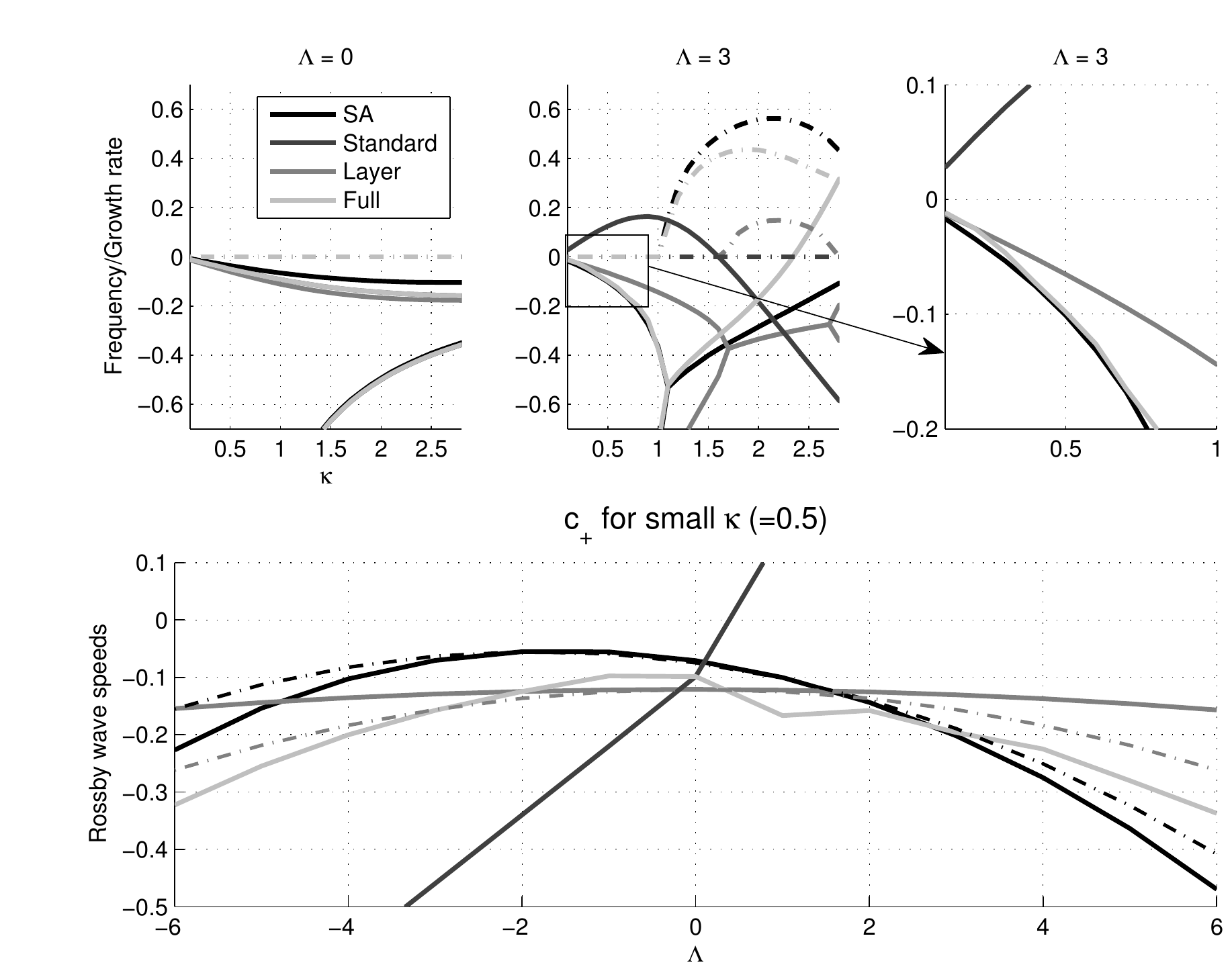}
\caption{Top: Comparison of the frequency (solid) and growth rate
  (dashed) from the two-SA-mode, two-standard-mode, two-layer model,
  and the full numerical solution for $\Lambda = 0$ (upper left) and
  $3$ (upper middle).  The panel on the upper right is a closeup of
  the stable branch at small wavenumbers for the case with $\Lambda =
  3$. Bottom panel: baroclinic Rossby wave speeds for the three
  truncated models at $\kappa=0.5$. Dashed lines are the
  small-$\kappa$ expansions for the analytic solutions given in
  \eqref{smallkappa}, up to $O(\kappa^2)$. The full solution (lightest
  gray line) is the curve on Fig.~\ref{fullxi} with $\xi = 0$ but
  without the absolute value or scaling factor $\pi^2$. The
  discontinuity in the full solution at $\Lambda = 1$ is because the
  system is stable when $0<\Lambda<1$ and changes to an eastward
  sheared Charney instability as $\Lambda$ increases. Note that, for
  the range of $\Lambda$ plotted, the small-$\kappa$ expansion for the
  analytical solution in standard modes overlaps with its numerical
  solution.}
\label{equal}
\end{figure}

The upper panels of Fig.~\ref{equal} plot the frequencies
$\Re(\omega)$ and growth rates $\Im(\omega)$ for each truncated model
(the analytical expressions are given in Appendix B), along side the
full numerical solution discussed in section 2, as functions of
wavenumber, for a few values of $\Lambda$.  With a non-zero surface buoyancy
gradient, the fact that the system generates faster westward Rossby
wave speeds (negative values of $\Re(\omega)$) for small wavenumber is
captured well by the SA modal solution. In addition, the growth rate
from the SA modal solution is also very close to that in the full
solution. The bottom panel of Fig.~\ref{equal} shows the baroclinic
wave speed for fixed wavenumber ($\kappa = 0.5$), as a function of
$\Lambda$, for all three truncated models as well as for the full
solution.  In addition, the small-$\kappa$ approximations in
\eqref{smallkappa} for the three truncated models are shown as dashed
lines.  The full solution, plotted here as the lightest gray line, is
the same as the curve on Fig.~\ref{fullxi} with $\xi = 0$, but without
the absolute value. The discontinuity of the full solution at $\Lambda
= 1$ is because the system is stable when $\Lambda$ is in between 0
and 1 and changes to eastward sheared Charney instability as $\Lambda$
increases.

The results in Fig.~\ref{equal} show that all solutions from the
truncated models are close to the full solution when $\Lambda=0$. In
the cases where surface buoyancy gradients are present, the SA-mode truncation is
closest to the true solution, in terms of both frequency and growth
rate. The standard mode truncation, however, essentially fails to
capture the dynamics. This may be because we have used the full
velocity (instead of the projecting the velocity into modes first) in
the wave equations \eqref{fqpsi}; but when $\Lambda\ne 0$, the
standard modes do not form a complete basis for $U$, resulting in
grave errors. The two-layer approximation does somewhat better at
large scales in the stable branch, while it completely underestimates
both frequency and growth rate at larger wavenumbers. Moreover, the
error of the layered model becomes larger in the next subsection when
more complicated systems are considered.

\subsection{Exponential stratification}

\begin{figure}[!htb]
\includegraphics[scale=0.9]{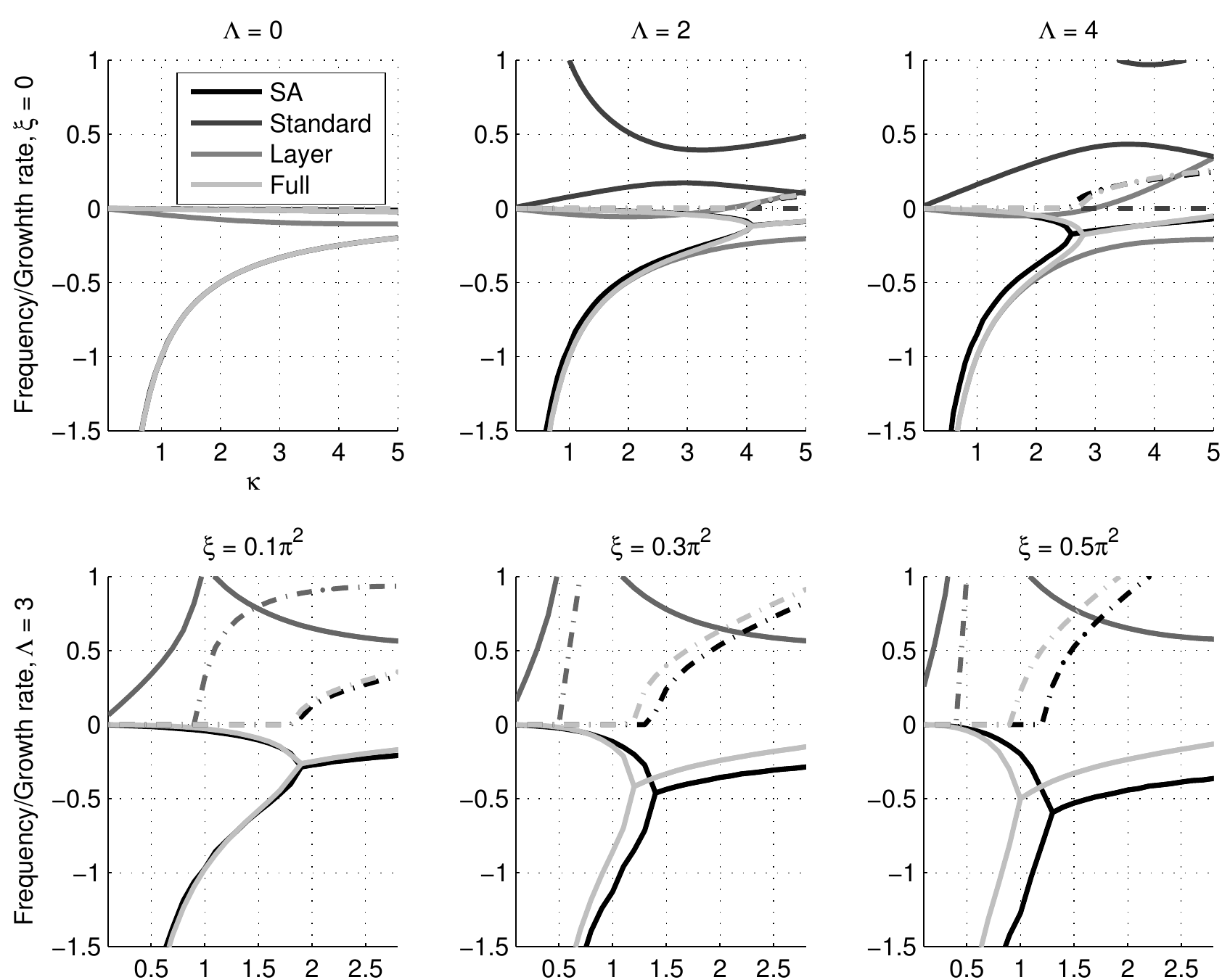}
\caption{Top panel: Comparison of frequency (solid) and growth rate
  (dashed) between the two-SA-mode truncation, two-standard-mode
  truncation, two-layer model and the full model, for varying
  $\Lambda$ with no interior flow ($\xi = 0$). Bottom panel:
  Comparison of frequency (solid) and growth rate (dashed) between the
  three-SA-mode truncation, three-standard-mode truncation, and the
  full model with fixed surface buoyancy gradient $\Lambda = 3$ and
  varying interior flows, $\xi = (0.1, 0.3, 0.5)\pi^2$.}
\label{expn}
\end{figure}

The interaction between surface buoyancy gradients and interior PV gradients in the
case of exponential stratification will be investigated with two types
of mean flows. We first consider the case with $\xi = 0$. When the
interior part of the mean velocity is absent, the corresponding
meridional mean PV gradient is $\Pi = 1 - \Lambda$, and therefore the
instability will be caused by the interaction between the surface
shear flow and the interior PV gradient as in section
\ref{constantn}. The system with this particular mean flow is
investigated with two-SA-mode, two-standard-mode, and two-layer
approximations. For the two-layer approximation, the depth of the
first layer is taken to be the fractional scale height $\delta$. The
top panel of Fig.~\ref{expn} gives a comparison of frequencies and
growth rates among the above three truncations for a few values of
$\Lambda$. From the dispersion relation, it can be shown that the SA-mode
solution and the two-layer approximation agree well with the full solution for
small $\kappa$. However, neither the two-layer approximation nor the
standard normal mode basis can capture the instability of the system,
while SA-mode solutions agree with the full solution, in both
frequency and the growth rate, for all scales.

The second type of mean flow has both surface and interior components,
i.e. $U= U^S + U^I$, thus the instability could be of the Charney-type
or Phillips-type, or both. For this case, with non-zero interior
velocity, three-mode truncations for both the SA and standard mode
bases are used (solutions are computed numerically), while the
low-resolution layered model is neglected. The bottom panel of
Fig.~\ref{expn} gives the comparison of those solutions with the full
solution for different combinations of $\Lambda$ and $\xi$. The
results show that, comparing to traditional baroclinic modal solution,
the SA-mode solution agrees well with the full solution, in both
frequency and growth rate.

\subsection{The efficiency of the surface-aware modes}

\begin{figure}[!htb]
\includegraphics[scale=0.9]{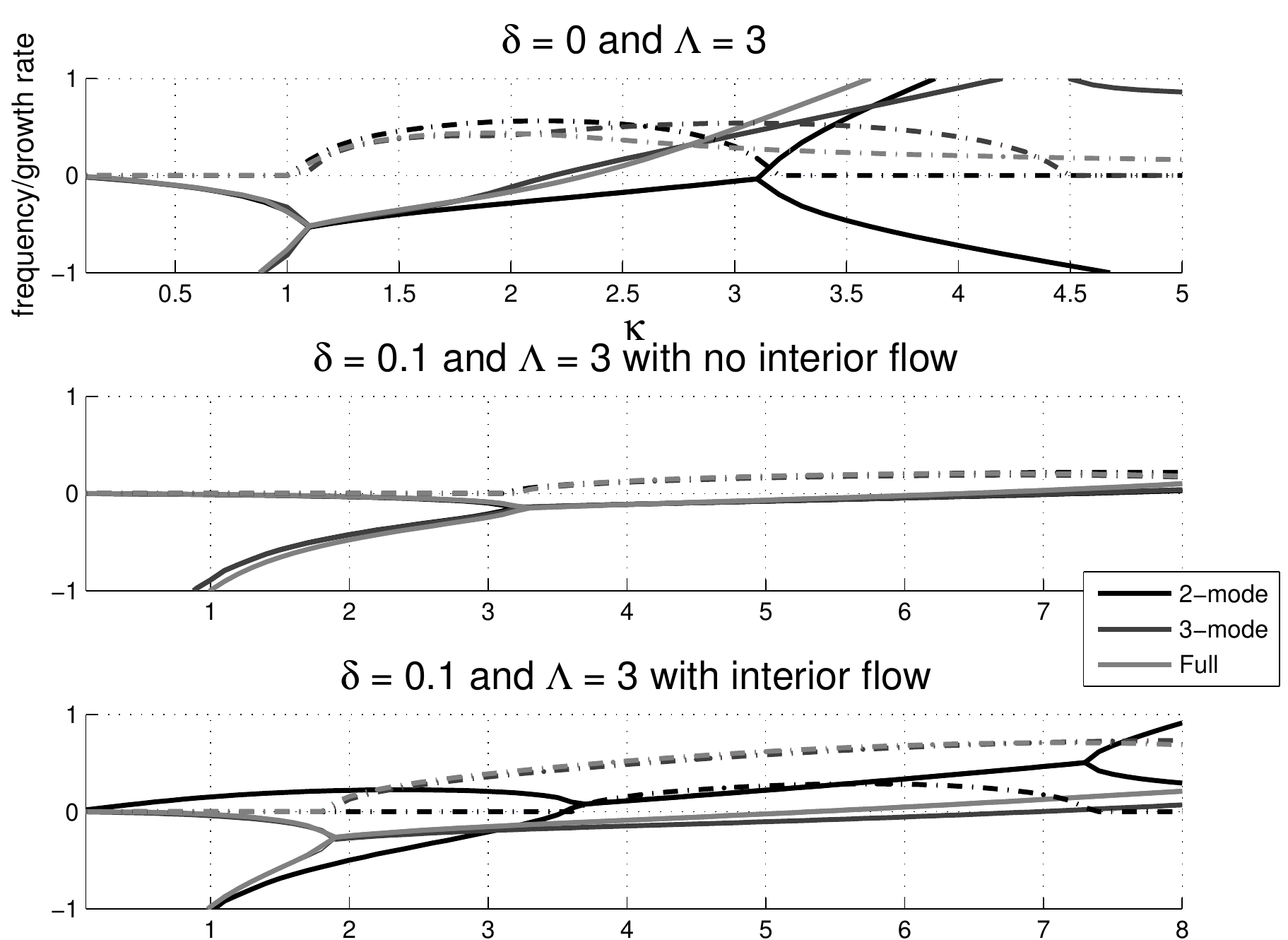}
\caption{The efficiency of the surface-aware modes: two-mode and
  three-mode SA-basis solutions, compared to full numerical solutions,
  for constant $N$ (Top), exponential $N$ without interior velocity
  (Middle), and exponential $N$ with interior velocity where
  $\xi=0.1\pi^2$ (Bottom).  }
\label{comp2m3m}
\end{figure}

A convergence study is preformed to test the efficiency of the SA
modal basis when surface buoyancy gradients exist. A comparison is made between
two-mode and three-mode truncated solutions in the SA-modal basis for
the three cases considered above, namely constant stratification,
exponential stratification without interior flow, and exponential
stratification with interior flow. Fig.~\ref{comp2m3m} gives the
comparison of frequencies and growth rates of two-mode and three-mode
truncations, along with the full numerical solutions, for these three
different combinations of mean flows and background
stratifications. The results show that SA modes capture the dynamics
efficiently with only a few modes. For all cases, the three-SA-mode
truncation is sufficient for solving this ocean Charney problem. On
the other hand, for the ocean Charney problem discussed here, one
should not expect solutions from standard normal modes to agree with
the full solutions; since this set of modes cannot form a complete
basis for systems with non-vanishing surface buoyancy gradients, uniform convergence
is lost if dynamical variables are expanded in these modes. 

\section{Conclusion and Discussion}\label{conclusion}

Based on our investigation using a simple mean flow
configuration to study Rossby wave speeds under the interaction
between surface buoyancy gradients and interior PV gradients, we find that in most
cases, surface buoyancy gradients yield faster Rossby waves.  However, one only
arrives at this conclusion when analyzing models that properly
represent lateral surface gradients. Since the ocean Charney problem
considered here has a non-vanishing boundary condition at the surface,
we expect the standard normal mode solutions, which assumes Neumann
boundary condition ($\phi_n' = 0$), not to match the full solution
well.

Three different truncated models are used to solve the ocean Charney
problem, namely a two-layer model, a truncated standard-mode model,
and a truncated SA-mode model. Among these three methods, only the
SA-mode model is derived from a non-standard Sturm-Liouville problem
with non-vanishing Neumann boundary conditions. Besides taking account
of lateral buoyancy gradients, the SA-mode truncated system also
provides a systematic way for computing wave speeds. If an $n$-mode
expansion is used, the phase speeds are the eigenvalues
of the $n\times n$ matrix formed by eigenmode expansion and
projection, as shown in equation \eqref{samatrix}. Taking advantage of
the simplicity of this algorithm, analytical solutions under the
two-mode truncation can be written explicitly. This is a fairly simple
mathematical model compared to the classic Charney problem
\citep{Pedlosky87}. Moreover, as shown by comparisons with full
solutions obtained by finite difference method, the wave dispersion
relation and growth rate can be well described by using only a few
modes.

One may notice that there are large discrepancies between the
truncated solutions in the standard normal mode basis and the full
solutions for all types of mean flows. These discrepancies are caused
by the fact that we directly used the mean flow in the wave
equations \eqref{fqpsi} when projecting those equations onto different
modes. One could possibly expect a better performance for the
solutions if one projected the mean flow onto the traditional
baroclinic modes first, then solved the wave equation. However, this
approach will eliminate the effects of surface buoyancy gradients on
the system since the projected mean flow does not have surface
buoyancy gradients. In contrast, since SA modes take into account the
information from the surface boundary, it is not surprising that
solutions from SA-mode truncation match the full solution. The layer
approximation, on the other hand, can be considered as the one which
captures the vertically-averaged dynamics of the system. Thus, a layered
model works well under linear background stratification because at
very large scale, solutions tend to represent the averaged dynamics.


The results suggest that, where lateral surface gradients are present,
a conceptual understanding of the wave propagation problems can be
obtained by using low-mode truncation in SA-mode basis. Interactions
between lateral surface buoyancy gradients and interior background PV gradients
exist in many regions in the ocean, and this fact leads to baroclinic
instability \citep{Tulloch2011}.  Analyzing the OCCA data used in
Fig.~\ref{drhody}, using local values of $f_0$, $\beta$ and $N_0$
(averaged over the upper 200 m), one finds $|\Lambda| \lesssim 2$ and
$\xi$ is $O(10^{-2})$, yielding wave speeds on order twice the long
Rossby wave speed, roughly in accord with satellite observations.  It
may also be worthwhile to apply this simple idea to data from 3D
simulations or direct ocean measurements to study surface buoyancy
effects on Rossby wave speeds and compare the results with
\citet{chelton96}.

\bigskip
\noindent\textbf{Acknowledgments:}  The authors wish to thank Ross
Tulloch for help with analyzing OCCA data; and Ian Grooms, Ed Gerber
and Glenn Flierl for helpful discussions.  This work was
supported by NSF OCE-0962054.  \bigskip

\section*{Appendix A: SA-mode construction}

Here we review SV13, with a focus on the ``ocean limit'' (with
vanishing lower surface buoyancy gradient) used in the present
work. The construction of the SA modes proceeds by simultaneous
diagonalization of two quadratic invariants for the system.  QG flow
conserves
\begin{align*}
  E_{\kappa} &= \frac{1}{2}\int_{-1}^0(s|\psi'|^2 + \kappa^2|\psi|^2)dz\\
  Z_{\kappa} &= \frac{1}{2}\int_{-1}^0 |q|^2 dz\\
  B_{\kappa} &= \frac{1}{2}|b|^2,
\end{align*}
which are the energy, potential enstrophy and upper surface buoyancy
variance, respectively.  In order to force the modes to represent
flows with surface buoyancy anomalies, one constructs a ``generalized enstrophy''
\begin{equation*}
  P_{\kappa} \equiv Z_{\kappa} + \alpha_+ B_{\kappa},
\end{equation*}
where $\alpha_+$ is an arbitrary nondimensional weight whose role
becomes more clear in the eigenvalue problem derived below.  To
proceed with the derivation, we define a a non-standard generalized PV
vector, an inner product, and two operators, as follows.  The
generalized PV vector is
\begin{equation}\label{Qdef}
  \bs{Q} = 
  \begin{bmatrix}
    b\\
    q(z)
  \end{bmatrix}= 
  \begin{bmatrix}
    \psi'(0)\\
    -\kappa^2\psi + (s\psi')'
  \end{bmatrix} 
\end{equation}
and the inner product is defined as
\begin{equation}\label{pd}
  \< \bs{Q}_1, \bs{Q}_2 \> = \int_{-1}^0 q_1^* q_2 ~\d z + b_1^* b_2,
\end{equation}
where $*$ denotes the complex conjugate.  The operators are
\begin{equation}\label{saops}
  \mathcal{E} Q = 
  \begin{bmatrix}
    \psi(0)\\
    -\psi(z)
  \end{bmatrix} 
  \quad\text{and}\quad
  \mathcal{P} Q = 
  \begin{bmatrix}
    \alpha_+ b\\
    q(z)
  \end{bmatrix}
\end{equation}
which are self-adjoint with respect to the inner product defined
above.  With these definitions, the invariants can be rewritten as
\begin{equation*}
  E_{\kappa} = \frac{1}{2}\< Q,\mathcal{E}Q \> 
  \quad\text{and}\quad
  P_{\kappa} = \frac{1}{2}\< Q,\mathcal{P}Q \>.
\end{equation*}

Demanding the simultaneous diagonalization of the quadratic forms
$E_{\kappa}$ and $P_{\kappa}$ is equivalent to solving the generalized
eigenvalue problem
\[
  \mathcal{P}\bs{\xi}_j = \mu_j^2 \mathcal{E}\bs{\xi}_j
\]
where the eigenfunctions $\bs{\xi}_j$ are analogous in structure to $\bs{Q}$.
Defining modes $\phi_j(z)$ analogous to the streamfunction $\psi(z)$
gives, by the operator definitions \eqref{saops},
\begin{equation}\label{saeval}
  \begin{bmatrix}
    \alpha_+ \xi_j(0)\\ \xi_j(z)
  \end{bmatrix}=\mu_j^2
  \begin{bmatrix}
    \phi_j(0)\\ -\phi_j(z)
  \end{bmatrix}.
\end{equation}
The eigenvectors $\bs{\xi}_j$ and $\phi_j$, as well as the eigenvalues
$\mu_j$ can be shown to be purely real.  The relationships
between $\xi_j(z)$, $\xi_j(0)$ and $\phi_j(z)$ are analogous to those
between $q$, $b$ and $\psi$ in \eqref{fqpsi}, respectively, allowing
the eigenvalue problem to be written entirely in terms of $\phi_j$ as
\begin{equation}\label{samodesfull}
  (s\phi_j')' -\kappa^2\phi_j = -\mu_j^2 \phi_j, 
  \quad\text{with}\quad
  \phi_j'(0) = \frac{\mu_j^2}{\alpha_+}\phi_j(0),
  \quad
  \phi_j'(-1) = 0.
\end{equation}
Note that in the limit $\alpha_+\rightarrow\infty$,
\eqref{samodesfull} becomes the Sturm-Liouville equation for standard
vertical modes \eqref{standardmodes}, which is obvious when $\mu_j$ is
eliminated in favor of $\lambda_j$ using the relation
$\mu_j^2=\kappa^2+\lambda_j^2$.  On the other hand, in the limit
$\alpha_+\ll 1$, the ``ocean-limit'' eigenvalue problem
\eqref{samodes} arises.  The eigenfunctions are orthogonal in the
sense that $\<\bs{\xi}_i,\mathcal{E}\bs{\xi}_j\>$ and
$\<\bs{\xi}_i,\mathcal{P}\bs{\xi}_j\>$ are both zero if and only if
$i\ne j$.  For our purposes, we choose the normalization
\begin{equation}\label{sanorm}
\<\bs{\xi}_i,\mathcal{E}\bs{\xi}_j\> = \int_{-1}^0 s \phi'_i\phi'_j+\kappa^2\phi_i\phi_j~\d z = \mu_j^2\delta_{ij},
\end{equation}
which is equivalent to \eqref{sanorm1} (but differs from that used in
SV13).  Given PV $q(z)$ and surface buoyancy $b$, one can construct
$\bs{Q}$ as in \eqref{Qdef}, and expand in the new modes as
\[
\bs{Q} = \sum_{j=0}^n a_j \bs{\xi}_j,
\]
and the coefficients $a_j$ can be recovered using \eqref{sanorm}, which yields
\[
a_j = \frac{1}{\mu_j^2}\left(\int_{-1}^0 s\phi_j'\psi' +
  \kappa^2\phi_j\psi  \right).
\]

Writing the wave equations \eqref{fqpsi} as a single equation in terms
of $\bs{Q}$ allows one to expand the equation in modes
\begin{equation}\label{saexp}
  \sum_{j=0}^n a_j \left[c ~\bs{\xi}_j - (\bs{U}+\bs{G})\circ \bs{\xi}_j\right] =0, 
  \quad\text{where}\quad
  \bs{U} \equiv \begin{bmatrix}U(0) \\ U(z)\end{bmatrix},
  \quad
  \bs{G} \equiv \begin{bmatrix}-\Lambda \\ \Pi(z)\end{bmatrix},
\end{equation}
and $\bs{A}\circ\bs{B}$ denotes the Hadamard product, or
element-by-element product, of two vectors $\bs{A}$ and $\bs{B}$.
In order to apply the orthogonality condition, we compute the
inner product $\mathcal{E}\bs{\xi}_i$ and \eqref{saexp} to get
\[
\sum_j a_j \left[c~\<\mathcal{E}\bs{\xi}_i,\bs{\xi}_j\> 
- \<\mathcal{E}\bs{\xi}_i,(\bs{U}+\bs{G})\circ \bs{\xi}_j\>\right] =0.
\]
By the self-adjoint property of the operator and the orthogonality
condition \eqref{sanorm}, the first inner product on the left is
$\mu_j^2\delta_{ij}$, and the other two terms are
\[
\<\mathcal{E}\bs{\xi}_i,\bs{G}\circ \bs{\xi}_j\> 
= \int_{-1}^0 -\phi_i \Pi \phi_j ~\d z - \phi_i(0) \Lambda \phi_j(0)
\]
and
\[
\<\mathcal{E}\bs{\xi}_i,\bs{U}\circ \bs{\xi}_j\> 
= \int_{-1}^0 -\phi_i U\left[(s'\phi_j')' - \kappa^2\phi_j\right] ~\d z 
+ \phi_i(0)U(0) \phi_j'(0).
\]
Using the eigenvalue problem \eqref{samodesfull}, the term in braces
inside the second integral can be replaced with $-\mu_j^2 \phi_j$.
Putting all the results together yields \eqref{samatrix} and
\eqref{Bmatrix}.  Note that one could also use the boundary condition
in \eqref{samodesfull} to replace $\phi_j'(0)$ with $(\mu_j^2/\alpha_+)
\phi_j(0)$ on the right hand side of the second expression, resulting
in the somewhat more succinct form 
\begin{equation}\label{Bmatrix2}
  \mathsf{B}_{ij} = \frac{1}{\mu_i^2}
  \left[\int_{-1}^0 \left(\mu_j^2U-\Pi\right)\phi_i\phi_j  ~ \d z 
   +  \left(\frac{\mu_j^2}{\alpha_+}U(0)-\Lambda\right)\phi_i(0)\phi_j(0)
  \right].
\end{equation}
However, this substitution doesn't work with $j>0$ in the ``ocean-limit''
used in the present analysis and so the form in \eqref{Bmatrix} is preferred.


\section*{Appendix B: Analytical solutions}

\subsection*{Expansion in two SA modes}

For the case with constant stratification with $U^I=0$, one
can use the expression for $U=U^S$ \eqref{usconstant}, and the ocean-limit
SA modes and eigenvalues in \eqref{saconstN} and \eqref{saevalconstN}
to compute the matrix $\mathsf{B}$ in \eqref{Bmatrix} (a number of
elementary but tedious integrations are needed).  With these in hand,
the matrix on the right hand side of \eqref{samatrix} reads
\begin{eqnarray}\label{twosa}
  \mathsf{B} =
  \begin{bmatrix}
    \frac{\Lambda-1}{\kappa\sinh(2\kappa)} 
    + \frac{\Lambda-1}{2\kappa^2} - \frac{\Lambda}{\kappa\tanh\kappa} + \frac{\Lambda}{3}
    &\pi\frac{2\Lambda( \mu_1^2\kappa\tanh\kappa-2\kappa^2)+ \mu_1^2}{\sqrt{2\alpha_+}\mu_1^4\kappa\tanh\kappa} \\ 
    \frac{\sqrt{2\alpha_+}\pi(1-\Lambda)}{2\mu_1^4}& 
    \frac{\Lambda-1}{\mu_1^4}-\frac{\Lambda}{\pi^2}  
  \end{bmatrix},
\end{eqnarray}
where $\mu_1^2 = \kappa^2+\pi^2/4$ was used to keep the
expresion more compact.  The eigenvalues can then be solved for
explicitly, and the result is
\begin{equation}\label{cphase}
c_{\pm} = \frac{1}{2} \Delta_3 \pm \frac{1}{2}\sqrt{\Delta_1^2 + \Delta_2}
\end{equation}
where
\begin{align*}
  \Delta_1 &= \mathsf{B}_{00} - \mathsf{B}_{11}\\
               &= \frac{\Lambda-1}{\kappa\sinh(2\kappa)} 
               + \frac{\Lambda-1}{2\kappa^2} 
               - \frac{\Lambda}{\kappa\tanh\kappa} 
               + \frac{\Lambda}{3} +\frac{1-\Lambda}{\mu_1^4}+\frac{\Lambda}{\pi^2}  \\
  \Delta_2 &=  4\mathsf{B}_{01}\mathsf{B}_{10}\\
	       &=2\pi^2 (1-\Lambda)\frac{2\Lambda( \mu_1^2\kappa\tanh\kappa-2\kappa^2)+ \mu_1^2}{\mu_1^8\kappa\tanh\kappa}
               \\
 \Delta_3 &=\mathsf{B}_{00} + \mathsf{B}_{11}\\
	      &= \frac{\Lambda-1}{\kappa\sinh(2\kappa)} 
              + \frac{\Lambda-1}{2\kappa^2} 
              - \frac{\Lambda}{\kappa\tanh\kappa} 
              + \frac{\Lambda}{3} +\frac{\Lambda-1}{\mu_1^4}-\frac{\Lambda}{\pi^2} 
\end{align*}
where $\mathsf{B}_{ij}$ are the elements of the matrix in \eqref{twosa}.
Notice that the phase speeds in equation \eqref{cphase} are independent
of the choice of $\alpha_+$ for the boundary condition of the eigenvalue
problem.

Expanding in a Taylor series about $\kappa=0$ gives
\begin{align*}
  c_{+} &= -0.08 -0.02\Lambda  
           + \left(0.009+0.01\Lambda-0.02 \Lambda^2\right)\kappa^2 + O(\kappa^4)\\
  c_{-} &= -\frac{1}{\kappa^2} + 0.005( 1 - \Lambda )+ O(\kappa^2).
\end{align*}

\subsection*{Expansion in two standard baroclinic modes}

In the limit $\alpha_+\rightarrow\infty$ the modal wave equation
\eqref{samatrix} becomes a problem for the coefficients of the
barotropic and baroclinic modes.  In this limit, $\phi_0 = 1$ and
$\phi_1 = \sqrt{2}\cos \pi z$, with eigenvalues $\mu_0^2 = \kappa^2$
and $\mu_1 = \kappa^2+\pi^2$ (since $\lambda_1 = \pi$ in our
nondimensionalization). The matrix $\mathsf{B}$ can again be computed
using \eqref{Bmatrix}, and the result is
\begin{equation}\label{2nmode}
  \mathsf{B} = 
  \begin{bmatrix}
    -\frac{1 -\Lambda}{\kappa^2} 
    & \Lambda\frac{\sqrt{2}}{\pi^2}\frac{\kappa^2 + \pi^2}{\kappa^2}\\
    \Lambda\frac{\sqrt{2}}{\pi^2}\frac{\kappa^2}{\kappa^2 + \pi^2} 
    & \frac{\Lambda}{4\pi^2} - \frac{1  - \Lambda}{\kappa^2 + \pi^2}
  \end{bmatrix}.
\end{equation}
This result can be compared to the same analysis in Salmon's textbook \citep{Salmon98}.
The eigenvalues are
\begin{align*}
  c_\pm = -\frac{1- \Lambda}{2\kappa^2} + 
  &\frac{(\kappa^2 + \pi^2)\frac{\Lambda}{4\pi^2} - (1 - \Lambda)}{2(\kappa^2 +  \pi^2)}\\
  & \pm \frac{1}{2}\sqrt{\bigg( -\frac{1-\Lambda}{\kappa^2} 
    - \frac{(\kappa^2 + \pi^2)\frac{\Lambda}{4\pi^2} 
      - (1 - \Lambda)}{\kappa^2 + \pi^2} \bigg)^2 + \frac{8\Lambda^2}{\pi^4}}
\end{align*}
Expanding about $\kappa=0$ gives the approximation
\begin{align}\label{nm2sol}
  c_+ &= 0.13\Lambda -0.025 
  + \left(0.02\frac{\Lambda^2}{1-\Lambda} - 0.01\Lambda  + 0.01\right)\kappa^2+ O(\kappa^2)\\
  c_- &= -\frac{1  - \Lambda}{\kappa^2} + O(\kappa^2).
\end{align}

\subsection*{Two-layer model}

The traditional truncation of the Rossby wave equation is
discretization into two isopycnal layers.  For the case of constant
stratification, the layers are taken to have equal depths, $H_1=H_2$,
and for the case of exponential stratification \eqref{Nexp}, we set
$H_1=\delta$ and $H_2=H-\delta$.  The discrete mean zonal velocities
$U_1$ and $U_2$ are taken to be vertical averages over the velocity
within each layer. The equations for for the streamfunction wave
amplitudes $\psi_1$ and $\psi_2$ for each layer are \citep{Pedlosky87}
\begin{equation}
(c - U_j)\bigg[ -\kappa^2 \psi_j- F_j(\psi_j - \psi_{3-j}) \bigg] + \Pi_j\psi_j = 0, \quad j=1,2,
\end{equation}
where, following our nondimensionalization, $F_j=2(H_1+H_2)/H_j$
(using Pedlosky's notation, $F_j = f_0^2 L^2/(g' H_j)$, where here
$N_0^2 = 2g'/H$, and our nondimensionalization length is $L=N_0
H/f_0$). The mean PV gradients are $\Pi_j = 1 - (-1)^jF_j(U_1-U_2)$.
For $U(z) = U^S(z)$, with $U^S$ given by \eqref{usconstant}, the mean
velocities are
\[
  U_1 = \frac{1}{2}\int_{-1/2}^0 U(z)~ \d z = \frac{\Lambda}{8} 
  \quad\text{and}\quad 
  U_2 = \frac{1}{2}\int_{-1}^{-1/2} U(z)~ \d z = -\frac{\Lambda}{8} = -U_1.
\]
For the case of exponential stratification, integrating $U^S$ in
\eqref{surfu} over the two layers gives
\[
U_1 = \frac{\Lambda}{6}(\delta^2 - 3\delta + 2)
\quad\text{and}\quad 
U_2 = -\frac{\Lambda}{6}\frac{\delta}{1-\delta}(\delta^2 - 3\delta + 2). 
\]

The wave speeds in the general case are
\begin{equation*}\label{layerc}
  c = U_2 + \frac{U_s\kappa^2(\kappa^2 + 2F_2) 
    - (2\kappa^2 + F_1 + F_2)}{2\kappa^2(\kappa^2 + F_1 + F_2)}
  \pm \frac{[(F_1 + F_2)^2 + 2 U_s \kappa^4(F_1 - F_2) 
    - \kappa^4U_s^2(4F_1F_2-\kappa^4)]^{1/2}}{2\kappa^2(\kappa^2 + F_1 + F_2)}
\end{equation*}
where $U_s = U_1-U_2$.  For constant background stratification
($F_1=F_2$, and $U_s = \Lambda/4$), these are
\begin{equation*}\label{layer2sol}
  c\pm = -\frac{\kappa^2 + 4}{\kappa^2(\kappa^2 + 8)} 
  \pm \frac{\bigg[ 64 - \kappa^4\Lambda^2(4-\frac{1}{16}\kappa^4) \bigg]^{1/2}}{2\kappa^2(\kappa^2+8)}.
\end{equation*}
A series expansion for small $\kappa$ yields
\begin{align}\label{layerk}
  c_+ &= -0.125+(-0.016\Lambda ^2 +0.016)\kappa^2+O(\kappa^4)\\
  c_- &= -\frac{1}{\kappa^2} + O(\kappa^2)
\end{align}

\section*{References}

\bibliography{lit}

\end{document}